\documentclass[10pt,preprint2,a4paper]{aastex}
\usepackage{graphics,epsf}
\usepackage{amsmath}                
\usepackage{amsfonts}               
\usepackage{amssymb}                
\usepackage{epsfig}                 

\def \kms{~\rm{km~s^{-1}}}

\def \yrs{~\rm{yrs}}

\def \Msun{~\rm{M_\odot}}
\def \Msyr{~\rm{M_\odot ~yr^{-1}}}

\begin{document}

\title{THE N II LINES IN ETA CARINAE: A FURTHER EVIDENCE FOR MASS TRANSFER DURING THE 19TH CENTURY GREAT ERUPTION}

\author{Amit Kashi\altaffilmark{1} and Noam Soker\altaffilmark{1}}

\altaffiltext{1}{Department of Physics, Technion $-$ Israel Institute of
Technology, Haifa 32000 Israel; kashia@physics.technion.ac.il;
soker@physics.technion.ac.il.}

\setlength{\columnsep}{1cm}

\begin{abstract}
We argue that the emission and the absorption components of he N~II lines recently
observed in $\eta$ Car can originate in the secondary's wind acceleration zone.
We base this claim in part on the presence of these lines in hot nitrogen rich stars, such as WN~8 stars,
and on the expected clumpy nature of such winds.
The envelope of the secondary star in $\eta$ Car is expected to be nitrogen rich
due to accretion of ${\rm few} \times \Msun$ of the primary luminous blue variable's nitrogen rich material during the
19$^{\rm th}$ century Great Eruption.
Another argument in support of N~II lines origin in the acceleration zone of the secondary wind
is the behavior of the emission components. The emission components of the lines show the same Doppler shift
variation as that of the absorption components.
The secondary's-wind origin of the N~II lines is compatible with the binary orientation in which the
secondary is closer to us near periastron passage.
\end{abstract}

\keywords{ (stars:) binaries: general$-$stars: mass loss$-$stars:
winds, outflows$-$stars: individual ($\eta$ Car)}

\small

\section{Introduction}
\label{sec:intro}

The origin of the visible He~I P~Cyg lines ($\lambda7065\rm{\AA}$, $\lambda5876\rm{\AA}$, and
$\lambda4471\rm{\AA}$) in the binary system $\eta$ Carinae is in dispute, with different researchers
attribute it to different regions, including the luminous blue variable (LBV) primary star or the winds collision region
(Falceta-Gon\c{c}alves et al. 2007; Humphreys et al. 2008).
In earlier studies (Kashi \& Soker 2007, 2008) we attribute the He~I P~Cyg lines to the
acceleration zone of the secondary's wind.
The Doppler shift of the P~Cyg absorption component is in agreement with the binary orientation with a longitude
angle $\omega=90^\circ$ (Kashi \& Soker 2008).
Namely, the secondary is closer to us at periastron passage.
This orientation, though in dispute, is strongly supported by numerous observations that were
discussed in details in Kashi \& Soker (2008, 2009b, 2011).

The main arguments for the He~I P~Cyg lines origin in the secondary wind's acceleration zone are as follows.
\begin{enumerate}
\item The Doppler shift of the P~Cyg absorption components follows very well the secondary's orbit (Kashi \& Soker 2007, 2008).
\item The lines are known to originate in stars even cooler than the secondary (e.g., Crowther \& Bohannan 1997).
\item The secondary luminosity that is $\sim 20\%$ of the system's luminosity, is sufficient to account for the depth of the He~I lines.
\item The Doppler shift of the emission follows that of the absorption. A model where the Doppler variation of the absorption comes from
changes in the radius where it forms within the primary wind, cannot explain the variation of the emission components of the line.
\end{enumerate}

Let us elaborate on the fourth argument.
Let us consider an alternative model where the He~I lines origin is the acceleration zone of the primary wind,
and the change in the Doppler shift of the absorbed part is attributed to a change in the
distance from the star where the lines are formed.
The Doppler shift changes as the wind velocity inside the acceleration zone is higher at larger distances from the star.
In such a model we would expect the width of the emission part of the line to change, but
its center to stay at about the primary stellar velocity.
These two properties are opposite to observations.
Observations show that the Doppler shift of the emission part, e.g., of He~I~$\lambda7067$,
follows that of the absorption, and its width does not change much (Nielsen et al. 2007).
This observed behavior is expected in a model where the entire source of the line changes its velocity,
as in a model where the line is formed in the secondary stellar wind.
The variation in the Doppler shift is attributed to the orbital motion of the secondary star.

In a recent paper, Mehner et al. (2011) present the $\lambda\lambda5668$--$5712\rm{\AA}$ lines in $\eta$ Car,
observed across the spectroscopic event of 2009.
Mehner et al. (2011) associate the lines with N~II, as previously classified by Hillier et al. (2001).
The Doppler shift variation with orbital phase of the N~II lines closely follows the behavior of the
He~I lines (Mehner et al. 2011).
Since N~II lines come from lower ionization regions, Mehner et al. (2011) conclude that neither
the He~I nor the N~II lines can originate in the secondary wind.
Mehner et al. (2011) do not provide a consistent explanation to the similar behavior of the N~II and He~I lines,
and to the Doppler variation with orbital phase.
They only comment that the observations of the N~II probably excludes some proposed models, such as those
where He~I lines originate in the secondary stellar wind or in an accretion disk.
We disagree with this conclusion of Mehner et al. (2011).
Below we argue that the behavior of the N~II lines might be accounted for in a model where the
N~II lines originate in the acceleration zone of he secondary wind.
Although even this model is not perfect, it is the only model that quantitatively accounts
for the variation of the Doppler shift.

\section{N~II lines in hot stars}
\label{sec:hotstars}

The effective temperature of the secondary is $T_{\rm eff} \simeq 37$--$40~\rm{kK}$
(e.g., Verner et al. 2005; Mehner et al. 2010).
Mehner et al. (2011) take it that this temperature is too high for the formation of the
$\lambda\lambda5668$--$5712\rm{\AA}$ N~II lines in the secondary's wind.

This conclusion possibly comes from the fact that observations of the $\lambda\lambda5668$--$5712\rm{\AA}$ N~II lines in
stars with such high effective temperatures are rare, and hard to find in the literature.
The natural candidates are WN star, as they are hot and nitrogen rich.
Herald et al. (2001) observed WR~40 and WR~16, classified as WN~8 stars with effective
temperatures of $T_{\rm eff}(\rm WR~40) \simeq 44 \rm{kK}$
and $T_{\rm eff}(\rm WR~16) \simeq 41.7 \rm{kK}$, respectively.

The presence of $\lambda\lambda5668$--$5712\rm{\AA}$ N~II lines in hotter stars than the secondary of $\eta$ Car
implies that if the wind of the secondary is nitrogen rich and dense, it might produce the N~II lines.

\section{Nitrogen-enriched secondary}
\label{sec:nitro}

The N~II $5668\rm{\AA}$ line profile (commonly called the N~II $5666\rm{\AA}$) in the left panel of figure 1 in Mehner et al. (2011),
shows an emission peak that maintains its width across periastron passage and
follows the Doppler shift variation of the absorption part.
This would not be the case if the emission of this line is an extended region that does
not orbit at high velocity around the center of mass.

Though the N~II and He~I lines intensities behave differently, their Doppler shift is the same.
This also shows that the entire lines-formation regions experience the same Doppler shift variation.
This cannot be accounted for in a model where the N~II and He~I lines-formation regions change their location within
the primary wind. A change in location within the secondary wind in our model
can take place, and it is even expected to occur, but it is not the main effect
(it causes deviation from pure orbital motion Doppler shift).

$\eta$ Car is a nitrogen rich system (e.g., Davidson et al. 1986; Dufour et al. 1999; Smith \& Morse 2004), and
it is accepted that the nitrogen rich gas is formed by the primary.
As we propose that the N~II lines originate in the secondary wind like the He~I lines,
the presence of nitrogen rich gas in the secondary wind requires an explanation.
The explanation sends us back to the event that made $\eta$ Car a famous star -- the $19^{\rm th}$ century Great Eruption (GE).
During the GE the LBV primary has lost $12$--$40~\rm{M_\odot}$ (Smith 2005; Smith \& Ferland 2007; Gomez et al. 2010).
Most of this mass was expelled to create the Homunculus nebula, but $\sim 3$--$6 \Msun$ from this nitrogen rich
material was accreted by the secondary (Kashi \& Soker 2011).
With the secondary mass loss rate of $\sim 10^{-5} \Msyr$ it would take the secondary
$\sim 4\times 10^5 \yrs$ to dissipate this mass.
According to the accretion model of the GE the present secondary wind is also nitrogen rich.

Nitrogen enrichment by itself is not sufficient in hot star.
Herald et al. (2001) could not reproduce in their model the N~II lines of the WN~8 stars WR~40 and WR~16.
They write:
``This problem may be alleviated by a more complex clumping implementation that allows for a
range of clump densities at each radius, as the N~II features would originate from the densest clumps.''
We take this claim, and suggest that the secondary wind is clumpy.
The clumps are pronounced in the acceleration zone where the velocity is $\sim 300$--$400 \kms$, the
same general region where we propose the He~I lines are formed in the secondary acceleration zone.
Probably, in that region some particular processes occur, such as instabilities and shocks,
with the possible formation turbulent regions.
Skinner et al. (2008), based on analysis of X-ray emission, suggest that shocks occur in
the acceleration zone of the wind of $\sigma$~Ori~AB (they could not resolve the two stars in X-ray),
where the wind velocity is much below the terminal speed.
Theoretically, instabilities, with the formation of clumps, in radiatively driven stellar winds can
develop deep in the acceleration zone where the wind is at much below its terminal speed
(Runacres \& Owocki 2002).

Over all, we suggest that where the wind of the secondary reaches a speed of $\sim 400 \kms$ in
its acceleration zone, instabilities that lead to shocks and the formation of clumps cause the
formation of some He~I and N~II lines.
Ferland (1999) showed that some lines can be enhanced in turbulent winds.
It is possible that the same occurs for the N~II lines in the wind of the secondary.

\section{Discussion and summary}
\label{sec:conclusions}

The Doppler shift variation of some He~I lines are quantitatively best explained by the orbital velocity of the
secondary star in $\eta$ Carinae (Kashi \& Soker 2007), for a semimajor orientation such that
the secondary is closer to us at periastron passage ($\omega=90^\circ$).
Other arguments that support this orientation are discussed in a previous paper (Kashi \& Soker 2008).
The attribution of the Doppler shift of the He~I lines to the secondary orbital motion explains the
observations that the emission components of the lines show the same Doppler variation as the
absorption components, a behavior that is in contradiction to models where the lines originate in the
primary wind.

Both the absorption and emission components of the N~II $5668\rm{\AA}$ line recently reported by
Mehner et al. (2011) show the same Doppler variation as the He~I lines observed by Nielsen et al. (2007).
For that, we argue here that the N~II $\lambda\lambda5668$--$5712\rm{\AA}$ complex, like some He~I lines,
originates in the acceleration zone of the secondary wind.

Mehner et al. (2011) {\it assume} that the N~II cannot be formed in the wind of the secondary of $\eta$ Car,
and from the similar Doppler shift variation with the He~I lines, come to the conclusion that also the
He~I lines do not originate in the secondary wind.
In this short paper we are strongly disputing this conclusion.

For the formation of the N~II lines in the hot secondary wind we require
($i$) that the secondary be nitrogen rich, and ($ii$) that dense clumps,
that have lower ionization parameter, are formed deep in the acceleration zone.
These requirements seemed to be fulfilled.
($i$) The more evolved primary star produces the nitrogen rich gas. The presence of nitrogen rich gas in the
outer layers of the secondary are the result of the accretion of $\sim 1 \Msun$ from the primary wind during the
1837--1856 Great Eruption of $\eta$ Car (Kashi \& Soker 2010).
($ii$) There are evidences, theoretically and observationally, that instabilities and
shocks, that form clumps, occur in the acceleration zone of winds of O stars, where the wind speed is much below its
terminal speed (e.g. Skinner et al. 2008).
Such clumps, for example, can account for the N~II lines in the hot WN~8 stars WR~16 and WR~40 (Herald et al. 2001).

We can summarize as follows.
Our claim that the N~II $\lambda\lambda5668$--$5712\rm{\AA}$ complex formed in the acceleration zone of the secondary
wind is compatible with other properties of the $\eta$ Car binary system.
\begin{enumerate}
\item The Doppler shifts of the N~II (reported by Mehner et al. 2011) and some He~I lines (reported by Nielsen et al. 2007)
are accounted for with the binary orbital motion.
\item The orientation required for the Doppler shift explains other properties, such as the X-ray absorption gas,
variation of narrow high emission lines, and more (Kashi \& Soker 2008).
\item The clumpy nature of the secondary wind claimed here, as well as of the primary wind argued for
by Moffat \& Corcoran (2009), facilitate the penetration of one wind through the other (after they are shocked).
Such clumps facilitate the accretion of the primary gas onto the secondary near periastron passages
(Akashi, Kashi, \& Soker, in preparation). The accretion process at each periastron passage seems to be
behind the spectroscopic event of $\eta$ Car (Kashi \& Soker 2009a).
\item The nitrogen enriched secondary wind is expected from the accretion model for the $19^{\rm th}$ Great Eruption
of $\eta$ Car. In addition, the accreted mass spun-up the secondary star. A rapid rotation can make the secondary wind
more complicated.
\end{enumerate}

\end{document}